\documentclass[12pt,preprint]{aastex}

\def\rel{{\rm rel}}
\def\e{{\rm E}}

\begin{document}
\title{A Second Method to Photometrically Align Multi-Site
Microlensing Light Curves:
Source Color in Planetary Event MOA-2007-BLG-192}

\author{
Andrew Gould\altaffilmark{1},
Subo Dong\altaffilmark{1,2,3},
David P.~Bennett\altaffilmark{4},
Ian A.~Bond\altaffilmark{5}, 
Andrzej Udalski\altaffilmark{6},
and
Szymon Kozlowski\altaffilmark{1},
}
\altaffiltext{1}
{Department of Astronomy, Ohio State University,
140 W.\ 18th Ave., Columbus, OH 43210, USA; 
gould,simkoz@astronomy.ohio-state.edu}
\altaffiltext{2}
{School of Natural Sciences, Institute for Advanced Study, Princeton, NJ,
08540, USA.  dong@ias.edui}
\altaffiltext{3}
{Sagan Fellow}
\altaffiltext{4}
{Department of Physics, Notre Dame University, Notre Dame, IN 46556, USA;
bennett@nd.edu}
\altaffiltext{5}
{Institute of Information and Mathematical Sciences, Massey University,
Private Bag 102-904, North Shore Mail Centre, Auckland, New Zealand;
i.a.bond@massey.ac.nz}
\altaffiltext{6}
{Warsaw University Observatory, Al.~Ujazdowskie~4, 00-478~Warszawa,Poland; 
udalski@astrouw.edu.pl}

\begin{abstract}

At present, microlensing light curves from different telescopes and
filters are
photometrically aligned by fitting them to a common model.  We
present a second method based on photometry of common field
stars.  If two spectral responses are similar (or the color
of the source is known) then this technique can resolve 
important ambiguities that frequently arise when predicting 
the future course of the event, and that occasionally persist
even when the event is over.  Or if the spectral responses are different,
it can be used to derive the color of the source when that is unknown.
We present the essential elements of this technique and apply
it to the case of MOA-2007-BLG-192, an important planetary event
for which the system  may be a terrestrial planet orbiting
a brown dwarf or very low mass star.  
The refined estimate of the source color that we 
derive here $V-I=2.36\pm 0.03$ will aid in making the estimate
of the lens mass more precise.

\end{abstract}

\keywords{gravitational lensing -- planetary systems -- methods numerical}

\section{{Introduction}
\label{sec:intro}}

The technical question most frequently asked of microlensers is
``How do you align different data sets''?  The standard answer
is that data are aligned via a common microlensing model.  That is,
there is a microlensing magnification model $A(t)$, and the
fluxes observed at the $i$th observatory are 
fit to $F_i(t) = f_{s,i} A(t) + f_{b,i}$,
where $f_{s,i}$ is the instrumental source flux for that observatory
and $f_{b,i}$ is blended light that does not participate in the event.
This approach is very powerful: it allows microlensers to work with
uncalibrated data, in non-standard filters, and to use difference
image analysis (DIA) photometry without worrying about the {\it flux}
zero point (which is simply absorbed into $f_b$).  These advantages
are all very important because microlensing data must often be
analyzed very quickly, even when they come from unexpected quarters.

Nevertheless, there are instances when an alternative alignment
method would be helpful.  As we outline below (and discuss more
extensively in Section~\ref{sec:discuss}) the need for such
an alternative method has been at least subconsciously apparent for
several years.  However, we were motivated to systematically
develop it by the problem of measuring the source color
for the interesting planetary event,  MOA-2007-BLG-192.

When microlensing planet searches were first proposed
\citep{liebes64,mao91,gouldloeb92}, there was no expectation that
the planet masses, distances, planet-star physical separations, or
orbital motion would be determined on an individual basis.  Rather,
it was thought that the quantities that could be measured were
the planet/star mass {\it ratio}, $q$, and the planet-star
projected separation $d$ in units of the Einstein radius, $\theta_\e$.
Physical information about the planets would be restricted to
statistical statements made about the ensemble of detections.

In fact, of the 9 microlensing planets published to date,
the masses, projected separations, and distances are measured for four
\citep{ob03235,bennett235,ob05071,ob05071b,ob06109},
and are at least partly constrained for the rest
\citep{ob05390,ob05169,mb07192,mb07400,mb08310}.
The primary reason for this turnabout is that,
in strong contrast to garden-variety microlensing events, planetary 
events usually give rise to measurable finite-source effects.
These then permit determination of 
\begin{equation}
\rho\equiv{\theta_*\over\theta_\e},
\label{eqn:rho}
\end{equation}
the ratio of the angular source size to the angular Einstein radius.
If $\theta_*$ can be determined, then one can measure $\theta_\e$
\begin{equation}
\theta_\e = \kappa M \pi_\rel,
\qquad \kappa \equiv {4 G \over c^2\,\rm AU}\sim 8.1\,{{\rm mas}\over M_\odot},
\label{eqn:thetae}
\end{equation}
where $M$ is the lens (host star) mass and $\pi_\rel$ is the 
lens-source relative parallax.  If one can then obtain a constraint
on another combination of lens mass and distance, from measuring
e.g., the so-called ``microlens parallax'' \citep{gould00}, the
flux from the lens \citep{han05,bennett07}, or astrometric offsets
\citep{bennett235,ob05071b}, then one can solve
for both $M$ and $\pi_\rel$, and so obtain the planet mass
(since $q$ is usually well-measured), as well as the distance
to the lens. (Since the source is almost certainly in the Galactic
bulge, $\pi_\rel$ directly yields the lens distance.)\ \ 
The lens distance, $D_{\rm L}$, then allows one to infer
the projected separation $r_\perp = D_{\rm L}\theta_{\rm E} d$.
Even if no other constraints are obtained, however, measurement
of $\theta_{\rm E}$ still yields the product $M\pi_\rel = \theta_\e^2/\kappa$,
which then provides statistical constraints on the properties
of the lens that are far better than if $\theta_{\rm E}$ is not
measured.

Hence, there is a high premium on measuring $\theta_\e$ during
planetary microlensing events.  The standard method 
for doing this is to measure the dereddened color $(V-I)_0$ and 
magnitude $I_0$ of the source during the event.  The dereddened
color gives the surface brightness \citep{kervella04}, and the
dereddened flux then gives the angular source size \citep{yoo04}.
In fact, to a good approximation, all one really needs is the
{\it instrumental} magnitude (which is automatically returned
by the light curve model) and the {\it instrumental} color
(which can be determined even without a model, just assuming
that one has near-simultaneous photometry in $V$ and $I$
at several different magnification levels.  The source can
then be placed on an instrumental color-magnitude diagram (CMD)
and compared to the position of the red giant clump, whose
dereddened color and magnitude are known fairly well.  Since
the source suffers nearly the same extinction as the clump,
one can directly determine $(V-I)_0$ and $I_0$ from such a diagram.

And therefore, microlensing planet hunters always try to obtain
$V$-band measurement while the source is significantly magnified,
to supplement the routinely-obtained $I$-band data.  
In fact, they try to obtain $H$-band data as well, since a
3-band $VIH$ determination can yield an even more precise
measurement of $\theta_*$ \citep{bennett10,ob07224}.  Nevertheless,
for a variety of reasons, including bad weather as well as the general
chaos that is an indelible part of chasing after high-magnification
microlensing events, sometimes these data are not taken or are
not of adequate quality.

In the case of MOA-2007-BLG-192, no $V$-band data were taken simply
because the event was not recognized as being sensitive to planets
until after peak, and was not recognized as containing a planet
until it had returned to baseline.  \citet{mb07192} were nevertheless
able to make a rough estimate of the source color by measuring
the source magnitude (as described above) and assuming that it
is a main-sequence star in the Galactic bulge.  While these assumptions
are not unreasonable, they lead to fairly large errors, and could
in principle fail catastrophically if the source happened, e.g., to be
in the Sagittarius Dwarf galaxy. Hence it would certainly be better
to have a measured color than an estimated one.  This is particularly
true because the planetary system detected in this event is quite 
interesting.  The most favored model is for a brown-dwarf host
with a few-Earth-mass planet.  Substantial work will be required
to obtain the necessary constraints to confirm or reject this model
\citep{mb07192}, but a color measurement (and so a measurement of
$\theta_*$) would certainly be a step in the right direction.

Here we present a general method of obtaining such post-facto
color measurements and apply it to MOA-2007-BLG-192.  We find
that it is somewhat redder than originally estimated, but well
within the previous (appropriately generous) error bar.

The method can potentially be applied to obtain colors of other
interesting microlensed sources.  Perhaps even more important, it can
be inverted to align data sets during the early phases of
microlensing events when the model is poorly constrained,
thus enabling much better real-time predictions, which are
crucial to organizing observations.  In some infrequent but
nonetheless important cases, the relative flux normalizations
from different observatories remain different for different
event models, even after the event is over.  Finally, it can
be applied to obtain ``microlens parallaxes'' by aligning
space-based and ground-based photometry.  The method
we describe here can be used to untangle all these cases as well.

{\section{General Method}
\label{sec:method}}

To measure the color of archival events, we 
take advantage of the fact the microlensing data
are often taken in non-standard bands.  For example, there
are many amateur observers who, because their telescopes
generally have small apertures, often obtain unfiltered data
or use very broad filters
\citep{ob05071,ob05169,ob07224,ob06109,mb08310,ob07050,ob08279}.
And, very importantly for the present case, the MOA collaboration
uses a broad $R/I$ filter, which we will refer to here as
$R_M$.  What one would like to do then, is to form an instrumental
CMD by combining photometry of a common set of field stars in two bands, 
the first being a standard
(or near-standard) $I$ band that is commonly used in microlensing
studies and the second being a non-standard band.  The source
fluxes are (as mentioned in Section~\ref{sec:intro}) routinely
returned by the model of the event, so the source position 
could be firmly located on this non-standard CMD.
Then one could identify the red-giant clump and measure the
offset of the source from the clump
(just as one does today in instrumental $V/I$ CMDs).  The remaining
step would be to make a color-color diagram that could relate the
offset so measured to the $V-I$ offset in standard Johnson-Cousins
bands.

In fact, as we will show in Section~\ref{sec:application}, such
an approach is not possible, or at least not optimal.  We adopt
a course that draws its inspiration from this approach but is
more flexible in dealing with several practical problems.

In the outline below, we will refer to the near-standard band
as $I_O$ and the non-standard band as $R_M$, but the reader
should keep in mind that the method can be used with any
two bands, whether standard or non-standard, provided only
that they have significantly different spectral response functions.
The method requires $3\times 2=6$ flux alignments or ``calibrations''.
[(1) Measurement of source flux in instrumental system.
(2) Calibration of field-star photometry relative to standard $V/I$.
(3) Alignment of source photometry and field-star photometry.]
$\times$ [(a) $R_M$.  (b) $I_O$]

Before describing how we apply this approach to MOA-2007-BLG-192,
it is worth reviewing how the same steps are ``taken care of''
in the more usual case when $V$ and $I$ photometry is obtained
during the event.  For step (1), the flux time series in the two
bands are fit to the microlensing model.  In fact, even if there
is no model, the color can be determined from a model-independent
regression of $V$ flux on $I$ flux.  This is often done, for example,
while the event is in progress and there is not yet a suitable
model.  Next, almost nothing
must be done for step (2), since the photometry is already in
standard (or near-standard) bands.  
Finally, step (3) usually requires no action at all.  If one uses
DoPHOT photometry \citep{dophot} for both the field stars and
the light curve, then these are automatically on the same
system.  Of course, it is common practice to model
light curves that are reduced using difference imaging analysis (DIA),
\citep{wozniak00,alard00}, which is generally superior to DoPHOT
for tracing the subtle details of planetary light curves.
However, DoPHOT is generally more than adequate for the much
coarser task of measuring the source color.  
Hence, in brief summary, for all
three steps, almost nothing needs to be done that would not be done anyway.
And this is perhaps the reason that it was not previously 
recognized that source colors could be obtained by combining
non-standard photometry.

{\section{Application to MOA-2007-BLG-192}
\label{sec:application}}

{\subsection{Instrumental Source Fluxes: DIA}
\label{sec:sourceflux}}

\begin{figure}
\plotone{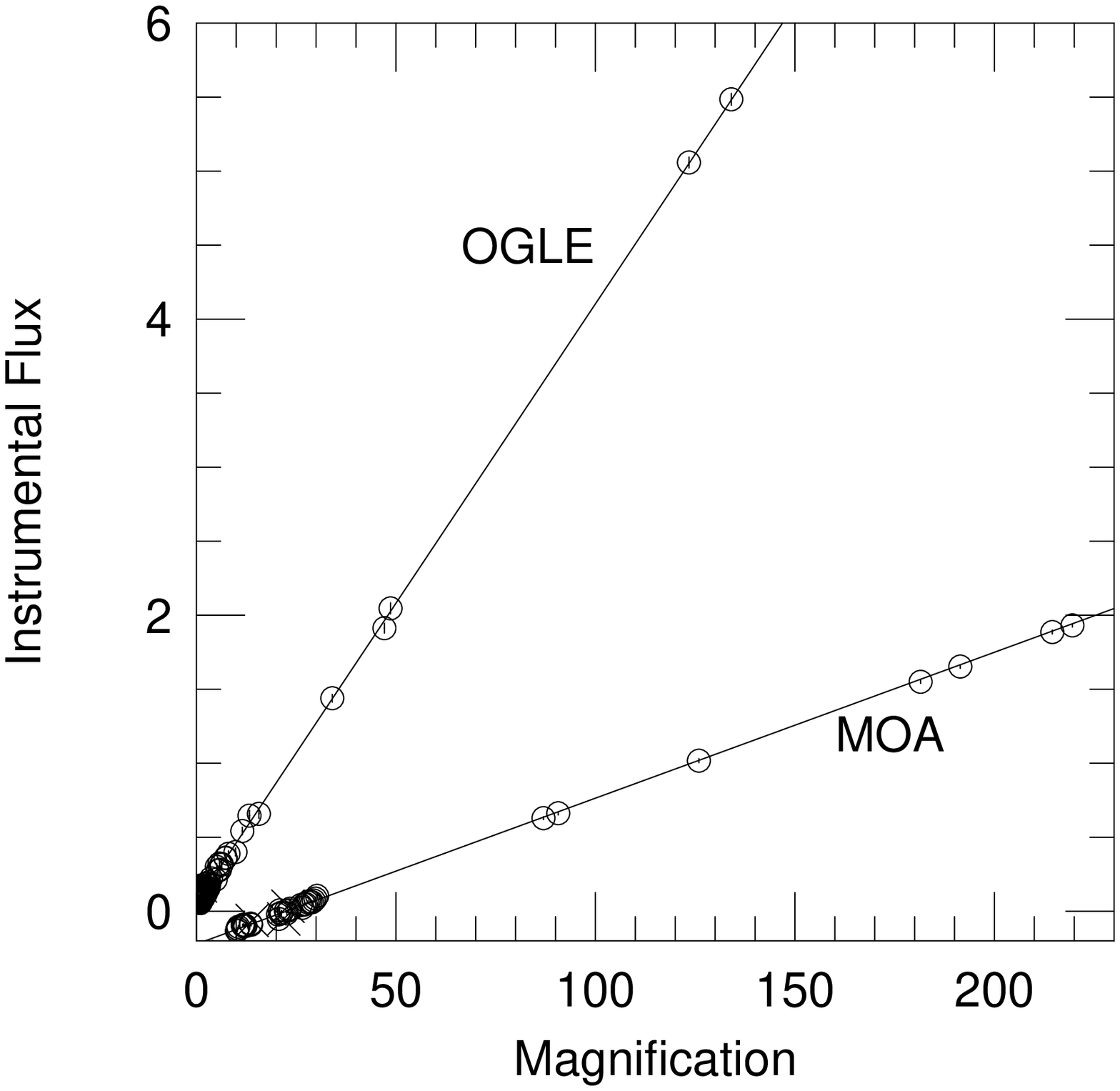}
\caption{\label{fig:fs}
Regression of instrumental fluxes from OGLE and MOA against the published
magnification model of \citet{mb07192}.  The slopes give the instrumental
source fluxes $f_s$.  Outliers beyond Gaussian expectation are recursively
rejected (crosses).  Remaining points (circles) have errors renormalized
to $\chi^2$/dof = 1.  The instrumental source fluxes $f_s$ are given
by the slopes, and the instrumental color is 
$(R_M - I_O) = -2.5*\log(f_{s,M}/f_{s,O}) = 1.5431\pm 0.0065$,
where $R_M$ and $I_O$ are the instrumental MOA and OGLE-III
passbands, respectively.
}
\end{figure}

As we show below, $\Delta(R_M - I_O)/\Delta(V-I)=0.265$.  Therefore,
any error in instrumental $(R_M-I_O)$ color (and so any error
in the individual source fluxes) will be multiplied by a factor
$\sim 4$ when we infer the $(V-I)$ color.  This implies that we must
attain maximum precision, which means using DIA rather than DoPHOT.
In principle this should not present any special problems since both
OGLE and MOA data are already reduced using DIA.  However, for reasons
described in Section~\ref{sec:calibration}, this does require that
we re-reduce the MOA data using software derived from
\citet{wozniak00} DIA rather than the \citet{bond01} 
version normally used by MOA.  Figure \ref{fig:fs}
shows the regression of measured (instrumental) source fluxes 
for the original OGLE data and the re-reduced MOA data, with respect to
the magnification $A$ of the published event model of \citet{mb07192}.

The slopes of the lines are the instrumental source fluxes $f_s$
because the observed difference flux is $f_{\rm obs} = A f_s + f_b$.
(Note that the {\it flux}
zero-point of this relation, $f_b$, plays no role in the
result.  This is important because difference imaging imposes an
arbitrary zero-point on the reported fluxes.)\ \
Outliers are recursively removed (crosses) if they exceed Gaussian
expectations, the errors of the remaining points (circles) are
renormalized to make $\chi^2$/dof = 1.  The imperceptibly small scatter
implies that the $f_s$ are very well determined (assuming that
the model is correct): $R_{s,M} = 23.0258\pm 0.0022$ and
$I_{s,O} = 21.4827\pm  0.0050$. 
Of course, the model is not
perfectly determined, so in practice the error in the source flux
is much larger.  However, changes in the model normally move the
source fluxes in tandem, so their {\it ratio} (and hence the
source color) does not depend strongly on the model.  If we
ignore all such model variation, we can combine the above measurements
of $f_s$ to obtain
$(R_M - I_O)_s = -2.5*\log(f_{s,M}/f_{s,O}) = 1.5431\pm 0.0060$.
To find the effect of model changes, we explore an ensemble
of models \citep{mb07192}
that all fit the data with $\Delta\chi^2\sim $ a few,
and find that the color dispersion [weighted by $\exp(-\chi^2/2)$]
is only 0.0025 (despite the fact that the dispersion in source
magnitudes is 0.045).  Adding this error in quadrature, we obtain,
\begin{equation}
(R_M - I_O)_s = -2.5*\log{f_{s,M}\over f_{s,O}} = 1.5431\pm 0.0065.
\label{eqn:rmis1}
\end{equation}

{\subsection{Calibration of Field-Star Photometry}
\label{sec:calibration}}

In this section, we align both MOA and OGLE-III photometry from the
event, with OGLE-II photometry \citep{szym05,udalski97}.  The latter
is calibrated, so in this sense we are ``calibrating'' these two
data sets.  However, that is not our primary objective.  Rather,
we are mainly using OGLE-II data to align these two data sets
with each other, and hence our primary focus is to carry out
the alignments with OGLE-II in as similar a manner as possible.
The main difficulty is that the MOA pixels are about twice as
large as OGLE pixels and the seeing is about 2.5 times larger.
Hence, our principal concern is that DoPHOT photometry of
MOA ``stars'' will, on average, include ``extra flux'' relative
to the corresponding OGLE-II stars, while OGLE-III stars will
not.  This would introduce a systematic error in the field-star
calibration that is not paralleled in the $f_s$ measurements
(which are done on difference images) and so would corrupt the
color measurement.

\begin{figure}
\plotone{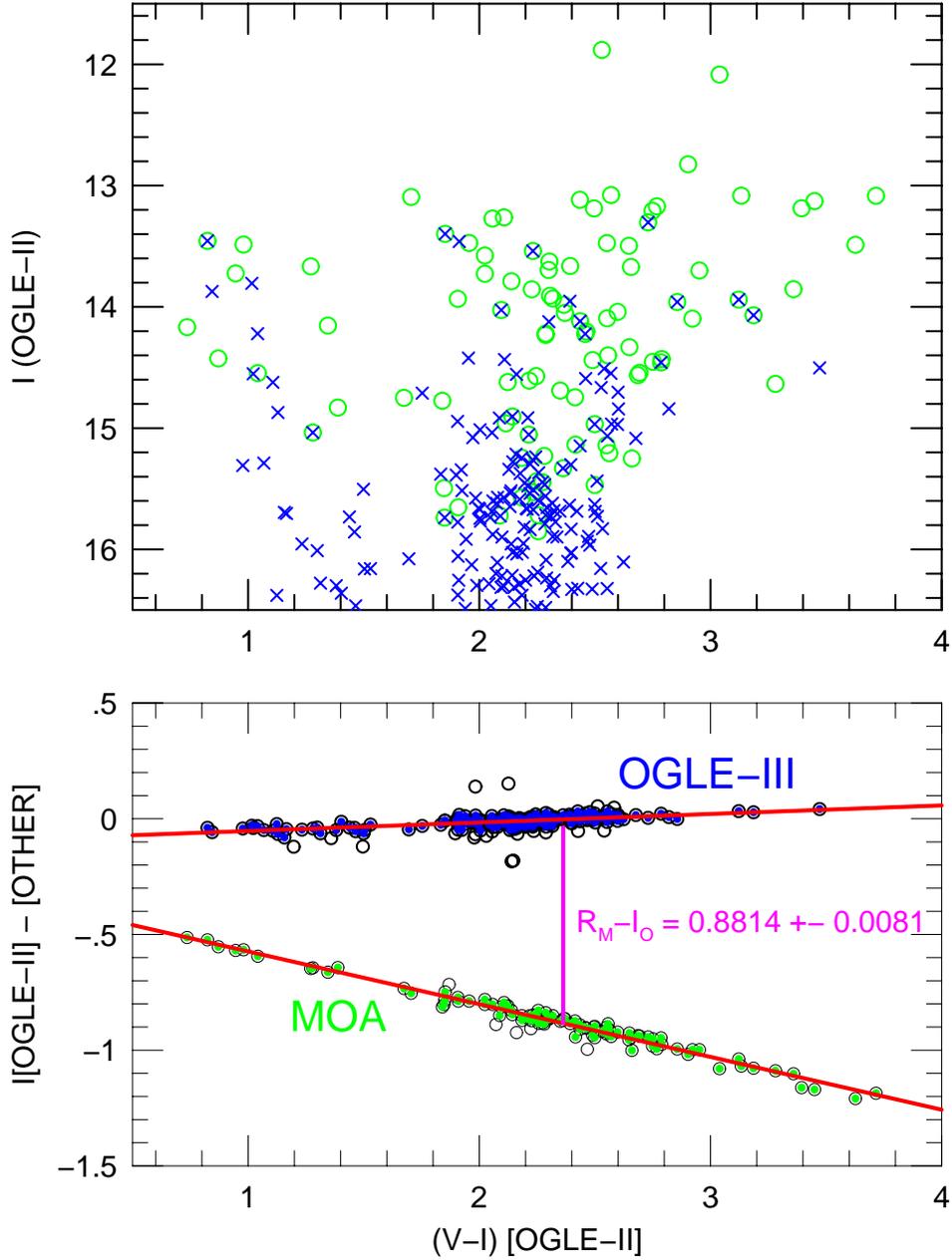}
\caption{\label{fig:calib}
Lower panel:
OGLE-III and MOA field stars are aligned with OGLE-II calibrated
photometry.  Only isolated stars are used for the comparison.
Then $2.5\,\sigma$ outliers are recursively rejected (open circles)
and remaining stars (filled circles) are fit to a straight line.
The difference is
$(R_M - I_O)_{\rm fld} = +0.8646\pm 0.0025 - (0.2648\pm 0.0044)[(V-I) - 2.3]$
meaning that the MOA-OGLE spectral baseline has only 26.5\% as
much leverage as $V-I$.
The vertical bar displays a geometric form of the final result.
Its indicated height is just
the sum of Eqs.~(\ref{eqn:rmis1}) and (\ref{eqn:align}).
When it is moved to the left until it is wedged in the ``jaws''
of the MOA and OGLE color-color relations, its horizontal
position gives the calibrated source color $(V-I)_s=2.363\pm 0.032$.
Upper panel is a CMD of the accepted stars for MOA (circles)
and OGLE-III (crosses).  
}
\end{figure}

To combat this difficulty we first construct a catalog of all
astrometric matches within $0.12''$, without regard to
magnitude offset.  Next, we consider all stars in the OGLE-II
catalog that lie within 3 FWHM of the matching catalog
(whether MOA or OGLE-III).  We compute the ratio of the
brightness of the wing of this potentially contaminating star 
to the central brightness of the target star.  If this ratio
exceeds 1\%, we exclude the target star from our sample.
Next we consider all stars within 1 FWHM of the target and
if any of these exceeds 2\% of the {\it total flux} of the
target, we also exclude the target.  In this way, we ensure
that the calibration is done only with isolated stars.
Finally, we fit to a function of the form,
$R_M = a + b(V-I)_{\rm OGLE-II}$, and recursively remove $2.5\,\sigma$
outliers.  We also remove the handful of stars with $V-I>4$ because
they are very far from our range of interest and have slightly
larger scatter (although this hardly affects the calculation).
The results for both MOA and OGLE-III are shown in
Figure~\ref{fig:calib}.  Numerically, 
$I - R_{M,\rm fld} = -0.8696\pm 0.0021 - (0.2280\pm 0.0035)[(V-I) - 2.3]$,
$I - I_{O,\rm fld} = -0.0050\pm 0.0013 + (0.0368\pm 0.0027)[(V-I) - 2.3]$, 
i.e.,
\begin{equation}
(R_M - I_O)_{\rm fld} = +0.8646\pm 0.0025 - (0.2648\pm 0.0044)[(V-I) - 2.3].
\label{eqn:calib}
\end{equation}

{\subsection{Alignment of Field-Star and Light-Curve photometry}
\label{sec:alignment}}

\begin{figure}
\plotone{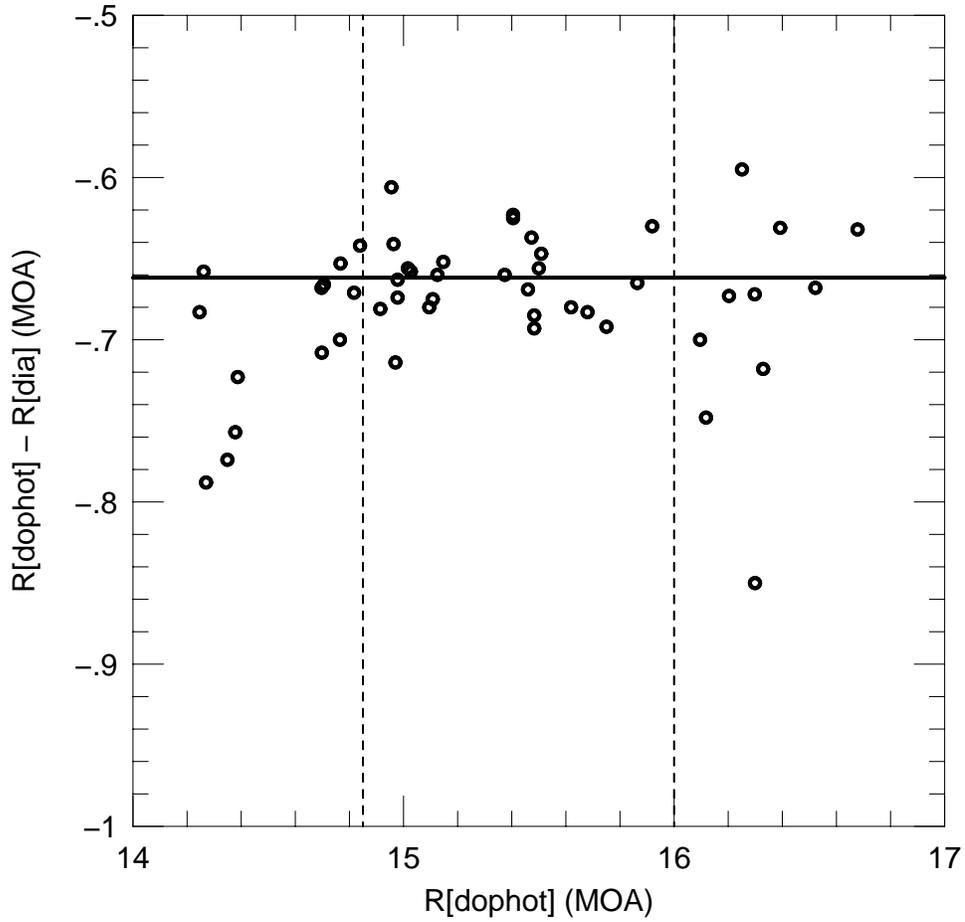}
\caption{\label{fig:align}
Offset between DoPHOT and DIA-PSF photometry for MOA data, the
former being designed for (unchanging) field stars and the latter
for difference images containing an (isolated) source.  The DIA PSF
can therefore only be applied to isolated field stars, which are shown
here.  Stars outside the dashed lines are too bright (and hence
non-linear) or too faint (and hence have too low signal) to be
included.  Offset is $R_{M,\rm DoPHOT} - R_{M,\rm DIA} = -0.6617 \pm 0.0049$.
}
\end{figure}

The MOA source flux was derived from the light curve in 
Section~\ref{sec:sourceflux} using DIA photometry, while the
field stars were calibrated in Section~\ref{sec:calibration}
using DoPHOT.  These must still be put on the same system.
In \citet{wozniak00} DIA, the difference images are photometered
using point-spread-function (PSF) fitting, and thus in principle
the same procedure can be applied to the field stars in the frame,
thus putting them on the same system.  However, DIA PSF fitting
is optimized in a very different way from DoPHOT PSF fitting.  On
the one hand, it must be able to measure negative fluxes (which DoPHOT
cannot), and on the other hand it is dealing with difference images,
which generally contain only variable stars and so are quite uncrowded.
In particular, therefore, DIA PSF fitting 
makes no attempt to deblend stars.  Thus,
it can only be applied to isolated stars.  Moreover, it appears
to be less robust than DoPHOT in dealing with mildly non-linear
to saturated pixels.  Hence, to transform from the DIA-PSF to
the DoPHOT system, one must make certain that comparison
is made only on isolated stars and avoids stars with 
mildly non-linear pixels.  We therefore begin by
restricting our comparison sample to isolated stars as described
in Section~\ref{sec:calibration}.  These are shown in 
Figure~\ref{fig:align}.  We exclude stars with $R <14.85$
because these have peak pixel values of 40,000 ADU, the point at
which the CCD becomes mildly non-linear \citep{moa3cam},
and we exclude those with $R>16$ to avoid
excessive scatter due to low signal. We find
\begin{equation}
(R_{M,\rm DoPHOT} - R_{M,\rm DIA})_{\rm fld} = -0.6617 \pm 0.0049.
\label{eqn:align}
\end{equation}
It appears visually from Figure~\ref{fig:align} that our non-linearity
threshold is sufficiently conservative, and we find
that if we are yet more conservative and exclude stars with $R<15$, the
result changes by $\ll 1\,\sigma$.

Note that the actual value of the offset $(-0.66)$ has no physical
meaning.  It is primarily the result of different normalization
conventions used by DoPHOT and DIA.  Secondarily, the DIA and DoPHOT
templates are different, the former being constructed by stacking
the best-seeing images, and the latter from an image in which
the source is highly magnified (and so easily recognized by the
DoPHOT software).

OGLE field stars are already on the DIA system, so no transformation
is necessary, i.e., $I_{O,\rm fld} = I_{O,s}$.  Hence, combining
Equations~(\ref{eqn:calib}) and (\ref{eqn:align}) yields
\begin{equation}
(R_M - I_O)_s = 1.5263\pm 0.0055 + (0.2648\pm 0.0044)[(V-I) - 2.3],
\label{eqn:rmis2}
\end{equation}
which combined with Equation~(\ref{eqn:rmis1}) yields 
\begin{equation}
(V-I)_s = 
 2.363 \pm 0.032 .
\label{eqn:rmis3}
\end{equation}

Figure~\ref{fig:calib} shows an alternative geometric derivation
of this result.  The height of the vertical bar is given
by the sum of Equations~(\ref{eqn:rmis1}) and (\ref{eqn:align}).
If it is moved to the left until it is wedged in the ``jaws''
comprising the MOA and OGLE color-color relations, its horizontal
position gives the calibrated source color $(V-I)_s$.

{\section{Test of Method}
\label{sec:test}}

We conduct a test of our method using a published event,
MOA-2008-BLG-310 \citep{mb08310} for which the source
$(V-I)$ color can be determined from $V$ and $I$ light curves
of the event.  We stress that we work strictly in 
{\it instrumental} magnitudes, whereas \citet{mb08310}
report results based on a calibrated version of the
same data.

First, in analogy to Equation (\ref{eqn:rmis1})
we fit the instrumental MOA DIA and CTIO DoPHOT lightcurve
fluxes to the magnifications found from the model
to derive $(R_{\rm M,DIA} - I_{\rm CTIO})_s =  0.6485 \pm 0.0088$, where
$R_{\rm M,DIA}$ is the instrumental MOA magnitude in the DIA system, 
and $I_{\rm CTIO}$ is the instrumental $I$-band magnitude in the
CTIO DoPHOT system.

Next, we match uncrowded stars from the DoPHOT and DIA templates,
and restrict consideration to the same flux range shown in 
Figure \ref{fig:align} to obtain
$(R_{\rm M,\rm DoPHOT} - R_{\rm M,\rm DIA})_{\rm fld} = -0.9625 \pm 0.0069$,
in analogy to Equation (\ref{eqn:align}).  Adding these two equations
yields $(R_{\rm M,DoPHOT} - I_{\rm CTIO})_s = -0.314 \pm 0.011$.

Next, we use field stars to make an instrumental color-color plot of 
$(R_{\rm M,DoPHOT} - I_{\rm CTIO})$ vs.\ $(V-I)_{\rm CTIO}$ and
find, in analogy to Equation (\ref{eqn:calib}),
$(R_{\rm M,DoPHOT} - I_{\rm CTIO})_{\rm fld} = -0.3140\pm 0.0030 + 
(0.155\pm 0.016)[(V-I)_{\rm CTIO} - 0.3]$.

Finally, we combine the previous two equations to predict
$(V-I)_{s,\rm CTIO,pred} = 0.300 \pm 0.071$.  This can be compared with
the instrumental color measured from the event light curve of
$(V-I)_{s,\rm CTIO,meas} = 0.310 \pm 0.011$.  This confirms, within
the relatively large measurement error, that the method works.

Note that the prediction is less accurate in this case than for
MOA-2007-BLG-192.  This is mostly due to the shorter color
baseline of $(R_{\rm MOA} - I_{\rm CTIO})$ relative to
$(R_{\rm MOA} - I_{\rm OGLE-III})$.  That is, CTIO $I$
is substantially bluer than OGLE-III $I$.

{\section{Discussion}
\label{sec:discuss}}

{\subsection{Implications for MOA-2007-BLG-192}
\label{sec:implications}}

The color measurement presented here, $(V-I)_s=2.36\pm 0.03$, is
0.13 mag redder than, but within the (justifiably generous) error bar of
$(V-I)_s=2.23\pm 0.20$
originally estimated by \citet{mb07192} based on the source apparent magnitude
$I_s=21.45$ and the assumption that the source was a typical
dwarf at the same distance as the observed clump stars.
This redder color by itself implies a 20\% lower surface brightness
and so a 10\% larger source radius, $\theta_*$.  However, we also make
several adjustments in the train of arguments that lead from $V/I$
measurements to $\theta_*$.  We begin by adopting the \citet{mb07192}
bulge clump color and absolute magnitude 
$[(V-I)_0,M_I]_{\rm cl}= (1.04,-0.25)$
and clump distance modulus 14.38, as well as their logic leading
to these values.  Hence, $[(V-I),I]_{0,\rm cl}= (1.04,14.13)$.
We remeasure the clump centroid on the CMD and
find $[(V-I),I]_{\rm cl} = (2.16,15.65)$.  Together, these
imply $[(V-I),I]_{0,s} = [(V-I),I]_s - [(V-I),I]_{\rm cl} +
 [(V-I),I]_{0,\rm cl} = (1.24,19.93)$.  Most importantly,
we use the very tight $VIK$ color-color relations of \citet{bessell88}
to infer $[(V-K),K]_0=(2.81,18.36)$ and then use the very tight
\citet{kervella04} $V/K$ surface-brightness relations to obtain
$\theta_*=0.57\,\mu$as.

To estimate the new error bar,
we first note that the error in the clump-offset method for estimating 
$(V-I)_0$
\citep{yoo04} has been determined to be 0.05 mag by direct
comparison with highly magnified dwarf stars using high-resolution
spectra (J.A.~Johnson, 2008 private communication).  This implies
$(V-I)_0 = 1.24\pm 0.06$, which by itself yields a fractional
error in $\theta_*$ of 3\%.  There are additional errors of
of 0.045 mag uncertainly in the model fit for the
source flux, of 0.04 mag in centroiding the height of the clump, as well
as much smaller errors in the \citet{bessell88} and
\citet{kervella04} relations.  Thus $\theta_*= 0.570 \pm 0.025\,\mu$as.
Finally, we state separately the error due to the assumed Galactocentric
distance $R_0= 8.0\pm 0.4$ kpc (since this may be resolved in the
relatively near future) and finally find $\theta_*= 0.57 \pm 0.04\,\mu$as,
compared to $\theta_*= 0.50 \pm 0.10\,\mu$as from \citet{mb07192}.
Hence, this color measurement essentially removes one of the important
uncertainties in characterizing the planet.

A key future test for the brown-dwarf hypothesis would be to
image the lens-source system using the adaptive optics on large 
telescopes or the {\it Hubble Space Telescope},
at various degrees of separation \citep{alcock01,kozlowski07}.
Because of the expected
faintness of the lens (regardless of the whether it is a brown
dwarf or a late M dwarf), independent knowledge of the source color
would be important in the interpretation of these images.

{\subsection{Application to Event Prediction}
\label{sec:application1}}

While we have presented our method in the context of measuring
the source color given a reasonably well-determined model, it
can easily be inverted to constrain models when traditional
methods of flux alignment fail.  The most common case is that
microlensing survey groups often notify the community of
newly discovered events and then go offline, either for short
periods due to daylight or for longer periods due to weather.
The events are then often monitored by other observers, but
the observations generally cannot be aligned with the discovery
data (with their long time baseline) using the traditional 
model-fitting technique because the models are completely
degenerate.  We have shown here that with good data, the
source fluxes, $f_s$, for different
data sets can be aligned to better than 1\%, provided that
the source color is known.  As mentioned in Section~\ref{sec:intro},
the color can be measured even without a model, and even if it
not measured, the alignment can be done from color estimates
provided that the spectral responses of the two instruments
are sufficiently close.  Of course, there will be uncertainties
in these alignments, but compared to the present situation
of complete ignorance, this would be a vast improvement
and would lead to greatly improved predictions of future
event behavior.  This
is especially important for high-magnification events, which
are the most sensitive to planets, and which often are not
discovered or not recognized to be high-magnification, until
a few hours before peak.  OGLE-2007-BLG-224 was an extreme
example of this \citep{ob07224}.

{\subsection{Application to Event Analysis}
\label{sec:application2}}

In some cases, different data sets cannot be aligned by
the traditional technique, even after the event is over.
For example, V.~Batista (2007 private communication) found
that MOA-2007-BLG-146 had two very different binary-lens solutions,
that differed strongly in their relative flux normalizations for
different observatories.
Another example is the planetary event OGLE-2005-BLG-071.
\citet{ob05071b} reported that there were data over one of the 
peaks from MDM and Palomar that could have helped constrain the
measurement of $\rho=\theta_*/\theta_\e$, but whose value was
substantially degraded because they could not be normalized
to other data sets.  And there are several planetary events
currently under analysis for which such flux degeneracies are
a significant obstacle to resolving model degeneracies.
The technique described here would be useful in all these cases.

{\subsection{Application to Space-Based Parallaxes}
\label{sec:application3}}

When \citet{gould95} proposed obtaining microlens parallaxes using
a single satellite, he argued that the spectral responses of
the space-based and ground-based cameras should be the same,
or at least that their differences should be known with extremely
high precision ($\Delta\bar\lambda \la 2\,$nm). This precision
requirement is rooted in the basic physics of the measurement: 
microlens parallax is derived from the difference in {\it magnifications}
as seen from two separated observers, but what is actually measured
is the difference in {\it fluxes}.  The observed flux is
given by $f = f_s A + f_b$, so to derive $A$ from $f$, one must know
$f_s$ and $f_b$, which depend on the overall microlensing model.
One can remove part of this ambiguity (namely $f_b$) by subtracting 
a baseline image $(A=1)$ from a magnified image.  Then one obtains
$\Delta f = (A-1)f_s$.  However, $f_s$ remains a fit parameter
for both observatories, which can have several percent errors, particularly
if (as expected) there are not many space-based measurements.  
However, \citet{gould95} argued, that if the spectral responses were
known to be the same, then even if the two $f_s$ were not measured
with great precision, the {\it ratio} of their values would move
in tandem, so that the {\it magnification difference} (needed for
the parallax measurement) would be known much better than the absolute
magnification.  The method of light-curve alignment presented here
can serve as a practical substitute for identical spectral responses
(which would be extremely difficult given that one telescope
is sitting below the Earth's atmosphere).

{\section{Conclusions}
\label{sec:conclude}}

We have presented a method for aligning microlensing 
light curves from different observatories that is 
independent of the standard method, which is based on fitting
to a common model.  We were initially motivated to develop
this technique in order to measure the source color for the
archival event MOA-2007-BLG-192, which is a candidate
brown-dwarf lens hosting a terrestrial planet.  We succeeded
in measuring this color within $\sigma(V-I) = 0.03$, which
will aid in future efforts to characterize this planetary system.

We have argued that the same technique potentially has much
broader uses, not only to find the colors of other source
stars, but also in the timely recognition of high-magnification
events and real-time analysis of anomalous events, which
are both critical to the data-gathering stage of microlensing
studies, as well as to the analysis of already-completed events.
Finally, we have shown that the technique can be used 
derive otherwise unobtainable microlens parallaxes by aligning
Earth-based and space-based lightcurves.


\acknowledgments

Work by A.G.\ and S.D.\ was supported in part by NSF grant AST 0757888.
Work by S.D.\ was performed [in part] under contract with the California
Institute of Technology (Caltech) funded by NASA through the Sagan
Fellowship Program.
I.A.B was supported by a grant from the Marsden Fund of NZ.
The OGLE project is partially supported by the Polish MNiSW grant
N20303032/4275 to A.U
We thank the MOA collaboration for making available the
images from MOA-2007-BLG-192 and MOA-2008-BLG-310.

\end{document}